\documentclass[a4paper,12pt, epsfig]{article}
\usepackage{epsfig}
\usepackage{amssymb}
\usepackage{amsfonts}
\usepackage{amsmath}
\usepackage{graphicx}
\newskip\humongous \humongous=0pt plus 1000pt minus 1000pt

\newif\ifdtup

\catcode`\@=11

\@addtoreset{equation}{section}
\def\theequation{\thesection.\arabic{equation}}

\def\@normalsize{\@setsize\normalsize{15pt}\xiipt\@xiipt
\abovedisplayskip 14pt plus3pt minus3pt%
\belowdisplayskip \abovedisplayskip
\abovedisplayshortskip \z@ plus3pt%
\belowdisplayshortskip 7pt plus3.5pt minus0pt}

\def\small{\@setsize\small{13.6pt}\xipt\@xipt
\abovedisplayskip 13pt plus3pt minus3pt%
\belowdisplayskip \abovedisplayskip
\abovedisplayshortskip \z@ plus3pt%
\belowdisplayshortskip 7pt plus3.5pt minus0pt
\def\@listi{\parsep 4.5pt plus 2pt minus 1pt
     \itemsep \parsep
     \topsep 9pt plus 3pt minus 3pt}}

\relax

\catcode`@=12

\catcode`\@=11

\def\section{\@startsection{section}{1}{\z@}{3.5ex plus 1ex minus
   .2ex}{2.3ex plus .2ex}{\large\bf}}

\def\thesection{\arabic{section}}
\def\thesubsection{\arabic{section}.\arabic{subsection}}

\def\appendix{\setcounter{section}{0}
 \def\thesection{Appendix \Alph{section}}
 \def\thesubsection{\Alph{section}.\arabic{subsection}}
 \def\theequation{\Alph{section}.\arabic{equation}}}
\def\SymBoxes#1#2#3#4{\newdimen\un@t \un@t#3%
\raisebox{#1}{\rule{#2\un@t}{#4}\hskip-#2\un@t% lower horizontal
\@tempdimb\un@t \advance\@tempdimb by-#4\@tempcntb#2\relax%
\@whilenum{\@tempcntb>0}\do{%                         % #2 vertical lines
\rule{#4}{\un@t}\hskip\@tempdimb \advance\@tempcntb by\m@ne}%
\hskip-#2\un@t \rule[\un@t]{#2\un@t}{#4}%
\rule[\un@t]{#4}{#4}\hskip-#4%             % upper horizontal line
\rule{#4}{\un@t}}\hskip-#4}                % rightest vertical line
\begin{document}
%\begin{letter}{~}

%%%%%%Define some new commands and  macros
\newcommand{\beq}{\begin{equation}}
\newcommand{\eeq}{\end{equation}}
\newcommand{\bea}{\begin{eqnarray}}
\newcommand{\eea}{\end{eqnarray}}
\newcommand{\beas}{\begin{eqnarray*}}
\newcommand{\eeas}{\end{eqnarray*}}
\newcommand{\defi}{\stackrel{\rm def}{=}}
\newcommand{\non}{\nonumber}
\newcommand{\bquo}{\begin{quote}}
\newcommand{\enqu}{\end{quote}}
%%%%%%%%%%%%%%%%
\renewcommand{\(}{\begin{equation}}
\renewcommand{\)}{\end{equation}}
%%%%%%%%%%%%%%%%%%%%%%%%%%%%%%%%%% definitions
\def \eqn#1#2{\begin{equation}#2\label{#1}\end{equation}}
\def\IZ{{\mathbb Z}}
\def\IR{{\mathbb R}}
\def\IC{{\mathbb C}}
\def\IQ{{\mathbb Q}}
\def\de{\partial}
\def\Tr{ \hbox{\rm Tr}}
\def\H{ \hbox{\rm H}}
\def\HE{ \hbox{$\rm H^{even}$}}
\def\HO{ \hbox{$\rm H^{odd}$}}
\def\K{ \hbox{\rm K}}
\def\Im{ \hbox{\rm Im}}
\def\Ker{ \hbox{\rm Ker}}
\def\const{\hbox {\rm const.}}
\def\o{\over}
\def\im{\hbox{\rm Im}}
\def\re{\hbox{\rm Re}}
\def\bra{\langle}\def\ket{\rangle}
\def\Arg{\hbox {\rm Arg}}
\def\Re{\hbox {\rm Re}}
\def\Im{\hbox {\rm Im}}
\def\exo{\hbox {\rm exp}}
\def\diag{\hbox{\rm diag}}
\def\longvert{{\rule[-2mm]{0.1mm}{7mm}}\,}
\def\a{\alpha}
\def\dag{{}^{\dagger}}
\def\tq{{\widetilde q}}
\def\p{{}^{\prime}}
\def\W{W}
\def\N{{\cal N}}
\def\hsp{,\hspace{.7cm}}
\newcommand{\C}{\ensuremath{\mathbb C}}
\newcommand{\Z}{\ensuremath{\mathbb Z}}
\newcommand{\R}{\ensuremath{\mathbb R}}
\newcommand{\rp}{\ensuremath{\mathbb {RP}}}
\newcommand{\cp}{\ensuremath{\mathbb {CP}}}
\newcommand{\vac}{\ensuremath{|0\rangle}}
\newcommand{\vact}{\ensuremath{|00\rangle}                    }
\newcommand{\oc}{\ensuremath{\overline{c}}}
\begin{titlepage}
\begin{flushright}
%ULB-TH/mm-dd\\
%hep-th/yymmnnn\\
\end{flushright}
%\bigskip
\def\thefootnote{\fnsymbol{footnote}}

\begin{center}
{\Large {\bf
Generalized BIons in M-theory\\
\vspace{0.3cm}
%Thermal Vacuum State
}}
\end{center}

\bigskip
\begin{center}
{\large  Chethan
KRISHNAN}\\
\end{center}

\renewcommand{\thefootnote}{\arabic{footnote}}

\begin{center}
%\vspace{0.2cm}
{\em  { SISSA,\\
Via Beirut 2-4, I-34014, Trieste, Italy\\
{\rm {\texttt{krishnan@sissa.it}}}\\}}

\end{center}

\noindent
\begin{center} {\bf Abstract} \end{center}
In string theory, stacks of D1-branes can expand into intersecting D3-branes. These configurations are called (generalized) BIons. We show how the analogous constructions in M-theory, where M2-branes blow up into calibrated intersections of M5-branes, arise from some of the %recently discovered 
membrane theories.

%\begin{center}
%{ {\footnotesize KEYWORDS}}: AdS-CFT
%correspondence, Black Holes, \\
%Quantum Field Theory in Curved Spacetime.
%\end{center}

%\begin{center}
%\vspace{1.6 cm}
\vfill

\end{titlepage}
\bigskip

\hfill{}
\bigskip

%\tableofcontents
\setcounter{footnote}{0}

BIons \cite{bion, const} are solutions of the BPS equations in the worldvolume theory of D1-branes, which correswpond to D1s expanding into D3s, or more generally, into intersecting configurations of D3s. In this note, we wish to see the emergence of analogous configurations in M-theory, where M2s expand into M5s. We will mostly work in the context of the membrane theory constructed by Bagger, Lambert and Gustavsson \cite{Bagger}. We are interested in BPS solutions of this theory which can be interpreted as fuzzy funnels of M-theory, known from the work of \cite{Basu}. To work up to it, we will first start with general arguments about the supersymmetries preserved by stacks of M2-branes\footnote{Due to space limitations, our citation list is minimal, the reader should consult \cite{KM} for references to original work. Membrane theories went through a revolution in the last months; references to more recent work can be found in, e.g., \cite{KM2}. Further aspects of fuzzy funnels in the context of ABJM theory \cite{ABJM} have been considered in \cite{new}.}\cite{review}.%, since these are ultimately what lead to the BPS equations. %The contents of this section are part of the general ``lore" in the field.

In the case of D-branes, the worldvolume theory is described by the transverse scalars, $X^I$, where $X^I$ are elevated to {\em matrices}. So for the case of M2's, we can start by trying to write a theory for $X^I_a$ where $I=(3,...,10)$ are the transverse directions and $a$ is a (multi-)index. %\footnote{It should be emphasized that we are using this merely as motivation. The index $a$ will turn out to be a 3-algebra index, and the representation theory fo 3-algebras is something we will not talk about at all.}.
From balancing various indices on either side, one can see that the most general (linear) way in which the 16 unbroken SUSY's can act is as
$\delta X^I_a=i \bar \epsilon \ \Gamma^I  \ \Psi_a, \ \  {\rm with} \  \ \epsilon=\Gamma_{012}\epsilon$.
If we assume canonical kinetic terms for the spinors and the scalars, in $2+1$ dimensions, we have
$[X]=\frac{1}{2} \  {\rm and} \  [\Psi_a]=1$,
as the scaling dimensions of the fields. With a bit of trial and error, it is easy to convince oneself that this means that the most general (without adding extra fields) SUSY variation that one can write down consistent with balancing spinor indices, internal indices and dimensions on either side is
$
\delta \Psi_a=\partial_\mu X^I_a \Gamma^{\mu I} \epsilon + c \ X^I_b X^J_c X^K_d f^{bcd}_{ \ \ \ \ a}\Gamma^{IJK} \epsilon
$
where $c$ is a parameter and the $f^{bcd}_{ \ \ \ \ a}$ are ``structure constants". The crucial observation of Bagger, Lambert and Gustavsson was to note that to close such a SUSY variation, one needs to covariantize the derivative  $\partial_\mu X^I_a$ in the above expression by introducing a gauge field. The rest follows more or less automatically upon demanding closure of SUSY on shell: {\bf (1)} The parameter $c$ gets fixed to $-\frac{1}{6}$, {\bf (2)} The structure constants $f^{bcd}_{ \ \ \ \ a}$ are to satisfy the so-called {\em fundamental identity}: $f^{[abc}_{ \ \ \ \ g}f^{e]fg}_{ \ \ \ \ d}=0$,
{\bf (3)} The equations of motion of the various fields are fixed.

The EOMs arising from the closure of the algebra can be obtained from an action. This is the BLG action (we will not write it down in full glory). But to construct that action, we need to assume two crucial things: {\bf (1)} The existence of a trace form $h^{ab}$ which can be used to raise indices so that we can construct scalars, {\bf (2)} $f^{abcd}\equiv h^{de}f^{abc}_{ \ \ \ \ e}$ is fully antisymmetric in all indices.
Unfortunately, if one restricts to positive definite $h^{ab}$, the only solutions to these restrictions is given by $f^{abcd}=\epsilon^{abcd}$. This choice is what corresponds to the original BLG theory. %We will say some words about indefinite $h^{ab}$ in the next section.

%\section{BPS Funnels}

To see BPS funnels in this theory, we will write down the scalar part of the BLG action:
$
{\cal L}_B=-\frac{1}{2}\Tr\,\Big(\partial_\mu X^I,\partial^\mu X^I\Big)
-\frac{1}{12}{\rm Tr}\Big([X^I,X^J,X^K],[X^I,X^J,X^K]\Big).
$
Here $X^I \equiv X^{I}_a\,T^a$, $[T^a,T^b,T^c] = f^{abc}{}_{d}\,T^d$, and $h^{ab}=\Tr\,(T^a,T^b)$ for the ``3-algebra" generators $T^a$. The BPS funnels arise when we set the energy functional computed from this Lagrangian to zero. This is because $Q |\psi\rangle =0$ implies $\langle \psi| \{Q, Q\} |\psi \rangle \sim \langle\psi| H|\psi \rangle =0$. Splitting off a total derivative piece from the Hamiltonian, we can write (for appropriate coefficients $g_{IJKL}$)
$
E=\frac{1}{2}\int d^2 \sigma \left( {\rm Tr}\left(\partial_\sigma
X^I-\frac{g_{IJKL}}{3!}
[X^J,X^K,X^L]\right)^2+T \right).
$
It is possible to write the Hamiltonian this way, if the configuration satisfies certain algebraic constraints in terms of the $X$'s\footnote{These constraints can also be viewed as arising from the consistency between the BPS equation and the equation of motion.}. For the calibrated intersections of M5-branes that the M2s can expand into, these constraints are automatically satisfied in BLG theory \cite{KM} due to the fundamental identity. It can also be checked that the other equations of motion arising in Bagger-Lambert, which where not visible in the {\em ad-hoc} constructions, are also satisfied\cite{KM}. Therefore we can consistently read off the first piece in the expression above as the fuzzy funnel equations of Basu-Harvey and Berman-Copland. Solutions of the BPS equation can be found by solving the auxiliary algebraic equation
$\frac{1}{6}g_{IJKL}[A^J,A^K,A^L]=A^I$,
because then $X^I(\sigma)=f(\sigma)A^I$ is a solution for $f(\sigma)$ satisfying $\partial_\sigma f(\sigma) = f^3(\sigma)$. The fuzzy funnels found in the literature can be constructed by a suitable definition of the 3-algebra using fuzzy 3-spheres \cite{Basu,KM}. Even though the structure is present, we point out that in BLG theory, the number of membranes is only two, so we need a more general theory of many membranes to have a complete picture of M-theory funnels.

One crucial ingredient in our Bogomolnyi positivity argument above is that it works only if the trace form $h^{ab}$ is positive definite, because otherwise the energy is unbounded below. This means that attempts to generalize BLG theory by relaxing this positivity \cite{ABC} do not fit into this picture. In negative trace form theories, the energy functional instead takes the form,
$
\mathcal{H} = \frac{1}{2}{\rm Tr}\Big(\partial_{\sigma}X^I \partial_{\sigma}X^I\Big) - \partial_{\sigma}X^I_+\partial_{\sigma}X_-^I + \frac{1}{12} {\rm Tr}\Big(X_+^I [ X^J, X^K] + ... \Big)^2.
$
The dots represent cycling the $I, J, K$ indices. This expression is written after expanding the 3-algebra expressions in terms of ordinary Lie algebras (from which the 3-algebras are constructed in these theories). In particular, the trace above is the usual trace of the Lie algebra and therefore positive definite. The negative trace of the 3-algebra gives rise to the negative sign of the $\partial_{\sigma}X^I_+\partial_{\sigma}X_-^I$ term. The $X_-^I$ is a Lagrange multiplier term enforcing the condition $\partial_\sigma^2 X^I_+=0$. If we solve for it by $X^I_+ \sim \sqrt{\lambda}$ with $\lambda > 0$, then the Hamiltonian is schematically that of a $\lambda \phi^4$ theory, and is positive definite. But the structure now looks like
$\ {\cal H} \sim (\partial_\sigma X + [X,X])^2\ $, which is suggestive of D2-D4 fuzzy 2-funnel intersections in Yang-Mills theory, whereas we need something like
$\ {\cal H} \sim (\partial_\sigma X + [X,X,X])^2\ $
to get fuzzy 3-funnels that connect M2s to M5s. In particular, we need three extra dimensions. We take this as further \cite{Mukhi} evidence that the negative trace form theories correspond merely to a rewriting of Yang-Mills theory.

\section*{\bf Acknowledgments}
I want to thank Carlo Maccaferri for the collaboration on which this contribution is based. This work was supported in part by IISN - Belgium (convention 4.4505.86).

% ==========================================================================
%
%%%%%%%%%%%%%%%%%%%%%%%%%%%%%%%%%%%%%%%%%%%%%%%%%%%%%%%%%%%%%%%%%%%%%%%%%%%%
%                      REFERENCES                            %
%%%%%%%%%%%%%%%%%%%%%%%%%%%%%%%%%%%%%%%%%%%%%%%%%%%%%%%%%%%%%%%%%%%%%%%%%%%%
%\newpage
%\bibliography{metasusy}

\end{document}